# Electron energy loss spectroscopy database synthesis and automation of core-loss edge recognition by deep-learning neural networks

Lingli Kong, Zhengran Ji, and Huolin L. Xin*

Department of Physics, University of California, Irvine, CA 92617, USA

Correspondence should be address to H.L.X. (huolin.xin@uci.edu)

**ABSTRACT**

The ionization edges encoded in the electron energy loss spectroscopy (EELS) spectra enable advanced material analysis including composition analyses and elemental quantifications. The development of the parallel EELS instrument and fast, sensitive detectors have greatly improved the acquisition speed of EELS spectra. However, the traditional way of core-loss edge recognition is experience based and human labor dependent, which limits the processing speed. So far, the low signal-noise ratio and the low jump ratio of the core-loss edges on the raw EELS spectra have been challenging for the automation of edge recognition. In this work, a convolutional-bidirectional long short-term memory neural network (CNN-BiLSTM) is proposed to automate the detection and elemental identification of core-loss edges from raw spectra. An EELS spectral database is synthesized by using our forward model to assist in the training and validation of the neural network. To make the synthesized spectra resemble the real spectra, we collected a large library of experimentally acquired EELS core edges. In synthesize the training library, the edges are modeled by fitting the multi-gaussian model to the real edges from experiments, and the noise and instrumental imperfectness are simulated and added. The well-trained CNN-BiLSTM network is tested against both the simulated spectra and real spectra collected from experiments. The high accuracy of the network, 94.9 %, proves that, without complicated preprocessing of the raw spectra, the proposed CNN-BiLSTM network achieves the automation of core-loss edge recognition for EELS spectra with high accuracy.

# INTRODUCTION

Ionization edges (core-loss edges) on the electron energy loss spectra (EELS) carry important information not only reflecting the elemental information but also allowing quantitative composition analysis. In addition, the edge onset shift and the near edge fine structure can be used to deduce the bonding state.[1] Compared to Energy Dispersive X-ray Spectroscopy (EDS), EELS can deliver electronic state information of a material and has higher sensitivity in detecting light elements.[2,3] However, the core-loss region of EELS is also featured by the low signal-noise ratio (SNR). Thus, precisely recognizing the elemental edges is the first step before any advanced analysis.

The traditional way of edge recognition involves multiple steps. Preprocessing operations, including background removal and sometimes multiple-scattering effect deconvolution, are conducted to extract the edges from the spectrum. Afterward, these edges are classified by referring to the database generated by verified experimental data.[4] Therefore, this process requires significant manual labor and is experience based.

Automatically finding ionization peaks in Energy Dispersive X-ray Spectroscopy (EDS) is now routine in commercial software because the peaks are distinctively above the background and the shapes are almost universally gaussian. However, automatically identifying edges in the EELS spectrum has been traditionally difficult because the edges are sitting on a large background, and in most cases, the jump ratio is low. Due to this nature, it has been non-trivial to locating EELS edges automatically. A number of approaches have been proposed to extract edges of unknown components in the spectrum, such as the spatial variant difference approach, multivariate statistical analysis[5] (independent components analysis (ICA),[6,7] principal components analysis (PCA)[8]), etc. Unfortunately, these approaches have constraints and are rarely used in existing EELS software because they are not always robust. Artificial intelligence (AI) has drawn the attention of researchers to simplify complex workflows and reduce labor work. Machine learning algorithms have been successfully implemented in mineral recognition and classification from Ramen spectra.[9] A dual Autoencoder-Classifier algorithm was also utilized to achieve the core-loss region denoising.[10] For detecting core-loss edges from the raw EELS spectra, Chatzidakis, et. al. proposed a convolutional neural network (CNN) for feature extraction and compound classification, but it was specifically trained for Mn element valance classification.[11] A network that can recognize all the elements on a raw spectrum remains to be developed.

Given that the ionization energy of a specific element is highly related to its electron configuration, nuclear charge, etc., its core-loss edge can only appear within a certain range along the energy loss axis. In other words, the core-loss edges of different elements appear almost "sequentially" on a spectrum. In addition, core-loss edges are present on the plasmon tail and the tails of lower energy core edges.[12,13] Thus, the edge shape is also energy loss dependent. As a result, the traditional convolutional neural network (CCN), which is originally designed for image classification, fails to classify the different core-loss edges on a raw spectrum. The reason being that a pure CCN ended with a fully connected classification layer that is only capable of classifying whether a certain feature is present in the signal/image; it, however, does not use the spatial information, although the information is encoded in the convolutional layers. For example, the latter transition metals, Mn, Fe, Co, Ni all have similar "rabbit-ear" shaped L2,3 edges. A pure CCN network would have the tendency to misclassify these edges because their locations on the energy axis are not used. That is why the region-based CNNs with localization/bounding box prediction layers have been developed to perform both classification and object localization.

In this respect, the recurrent neural networks (RNNs) utilize the correlation in the "time" domain and have been proven to be capable of handling sequence prediction problems. Based on this consideration, we built a Convolutional-Bidirectional Long Short-Term Memory (CNN-BiLSTM) network by combining a CNN network with an RNN network, i.e., the Bidirectional Long Short-Term Memory (BiLSTM) network. Given sufficient training, the model is capable of achieving the automation of ionization edge recognition on raw EELS spectra. To build a large and diverse ground truth labeled training dataset, we digitized a large number of core-loss edges collected from real experiments and literature; we built the ground truth-labelled spectrum database by synthesizing spectra by including experimental conditions, including the multiple scattering effect, the uncertainty of the energy axis due to stray field and hysteresis of the prism, the low-frequency patterned noise, etc. Our well-trained CNN-BiLSTM network was tested against both the validation spectra and the real spectra collected from experiments, with average F-1 scores of 0.98 and 0.99, respectively. For the real spectra dataset prediction, only 1.1 % of the predicted spectra have false-negative elements, and 94.9 % of them are accurately predicted where the predicted edges exactly match the ground truth. The sensitivity of the network was further evaluated by detecting the lowest signal noise ratio at the edge peak (PSNR) for each element, at which the network can recognize the edge. Correspondingly, the detectable jump ratios were obtained. Based on our observation, the detectable jump ratios and PSNR are very low meaning that our networks are accurate in low-dose situations. It suggests that our network is remarkably sensitive to the edge signal. The high accuracy and sensitivity demonstrate that our network is highly reliable for core-loss edges recognition.

**METHODS**

**Database Construction**

Considering that the available spectra from our own experiments are not enough for training a deep learning model, we built a spectrum database containing 250,000 simulated spectra, which are synthesized based on real ionization edges obtained from experiments/literature. The various physical phenomena in experiments are embedded into the synthesized spectra by convolving the broadening effects and simulating the instability with randomness. The simulated spectra are in close resemblance to the real EELS spectra and are suitable for our CNN-BiLSTM network training.

*Edge Modeling* We digitized around 400 core-loss edges covering 20 different elements from journal papers. The edges are then fitted by using the multi-gaussian model to locate the coordinates of the peaks and smooth out the noise effects as shown in Fig. 1(a).[13] Fig. 1(b) presents some examples of the fitted edges for different elements. We can notice that the fitted curves round off the noise fluctuations well with the near edge fine features well preserved.

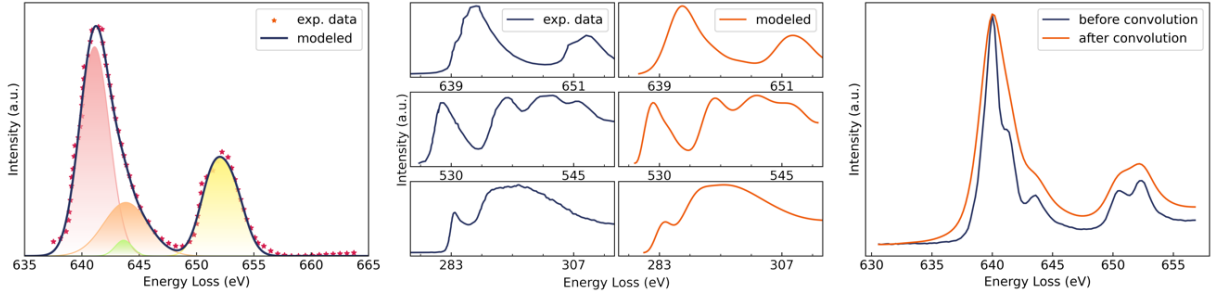

(a) multi-gaussian fitting of Mn-L edge  (b) original and modeled edges   (c) Mn-L edge with convolution

Fig. 1 Ionization Edge Modeling

In the real case, when a spectrum is recorded by the detector, the instrument broadening is also convolved in it. Therefore, we further processed the modeled edges by convolving them with the corresponding broadening function, i.e., the point spread function (PSF). The intensity with scattering effect can be expressed as $I_r$:

$$I_r(E) = I_u(E) \otimes \text{PSF}_{\text{ZLP}}$$

where $I_u$ is the original intensity without the broadening effect and $E$ is the energy loss.

The scattering effect is composed of amonochromaticity of the beam and broadening due to the detectors. Thus, PSF is constructed by two terms:

$$PSF_{ZLP} \propto PSF_{amono} \otimes PSF_{detector}$$

where $PSF_{amono}$ represents the energy spread of the incoming electron beam, and $PSF_{detector}$ represents the broadening in the detectors. For the amonochromaticity effect, the form of the Gaussian distribution function is used:

$$PSF_{amono} \propto e^{-\frac{E^2}{2\sigma^2}}$$

where $\sigma$ is the standard deviation, and can be expressed by:

$$\sigma = \frac{FWHM_g}{2\sqrt{2 \ln 2}}$$

$FWHM_g$ is the full-width half-maximum of the Gaussian distribution. In our data augmentation, it ranges from 0.1 to 1.6.

For the detector broadening effect, the Lorentzian distribution function is commonly used to describe the diffused tail (caused by light diffusion in the scintillator and the optical fiber coupler), which is expressed by:

$$PSF_{detector} \propto \frac{HWHM_l}{E^2 + HWHM_l^2}$$

$HWHM_l$ is the full width at half maximum of the Lorentzian distribution. In our data augmentation, it ranges from 0.1 to 0.6.

With the broadening effects convolved, the synthesized edges will be comparable to those obtained from the different instruments. Fig. 1(c) shows the Mn-L edge before and after convolving the PSF. We can observe that the broadening effect of the point spread functions (PSFs) has been successfully embedded into the modeled edge. It is worth mentioning that the Lorentzian broadening is often neglected by researchers in the field but it is a critical broadening effect that shall be included to reproduce an experimental EELS spectrum.

*Background and Jump Ratio*   In the core-loss region of the spectrum, it has been commonly accepted that the background can be modeled by using the inverse power law equation:

$$J(E) \propto E^{-r}$$

where *r* can vary across the specimen as a result of changes in thickness and composition. Typically, the value of *r* fluctuates around 1.88. In our case, we assign the value based on a normal distribution with a mean of 1.88 and a standard deviation of 0.5. The values less than 0.1 in the normal distribution are cut off to guarantee that the background remains realistic.

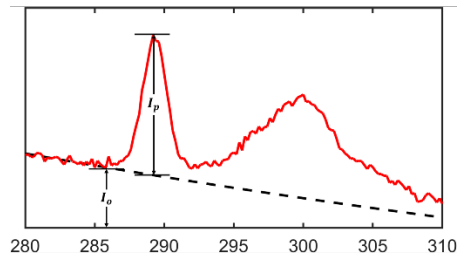

Fig. 2 Jump Ratio and Peak Signal Noise Ratio

To embed the modeled edges into the background, the jump ratios of the edges must be assigned. The jump ratio is defined as the ratio of the intensity difference between the peak intensity and the background $I_p$ and the intensity of the edge start point $I_o$, which is shown in Fig. 2. In our synthesized spectrum, the minimum jump ratio of the edges is set to 0.15. Because using data with too low a SNR for training would result in overfitting and high false positive rate, we choose a noticeable minimum jump ratio (>0.15) to insure low false positive rate. Based on the digitized experimental data, we use the maximum jump ratios from experimental spectrum to set the upper limit of the jump ratios for the corresponding element. We summarized the edges of 20 elements collected from the literature and the jump ratio range in Table 1.

Table 1 Edge Library[14-65]

| Element | Jump Ratio Range | Number of Edges | Element | Jump Ratio Range | Number of Edges |
|---|---|---|---|---|---|
| B | 0.15~0.5 | 31 | Mn | 0.15~0.5 | 26 |
| C | 0.15~1.5 | 25 | Fe | 0.15~1.5 | 39 |
| N | 0.15~0.7 | 14 | Co | 0.15~0.7 | 23 |
| O | 0.15~1.5 | 38 | Ni | 0.15~1.5 | 20 |
| F | 0.15~1.0 | 8 | Cu | 0.15~1.0 | 17 |
| Ca | 0.15~1.3 | 13 | Si | 0.15~1.3 | 34 |
| Sc | 0.15~1.3 | 6 | P | 0.15~1.3 | 5 |
| Ti | 0.15~1.5 | 47 | S | 0.15~1.5 | 1 |
| V | 0.15~1.5 | 40 | Cl | 0.15~1.5 | 1 |
| Cr | 0.15~1.5 | 18 | Ar | 0.15~1.5 | 1 |

***Spectrum Synthesize and Data Augmentation***    The spectrum synthesis process is summarized in Fig. 3. The first step is to randomly pick 1 to 6 edges from the edge library and convolve them with the PSFs. Fig. 3 (a) shows the 6 convolved edges we used in the example. Fig. 3 (b) lists the augmentation operations before obtaining the final spectrum shown in Fig. 3 (c).

Since we only consider the core-loss region on spectra, the energy loss range is set from 50 eV to 1100 eV. The background $J(E)$ is normalized to $J_n(E)$:

$$J_n(E) = \frac{E^{-r}}{50^{-r}}$$

Before adding the edges to the background, to simulate the long tail of the core-loss edges and obtain a smooth merged spectrum, we extend the edges using another inverse power law, where the range of the power is from -3.8 to -1.8.

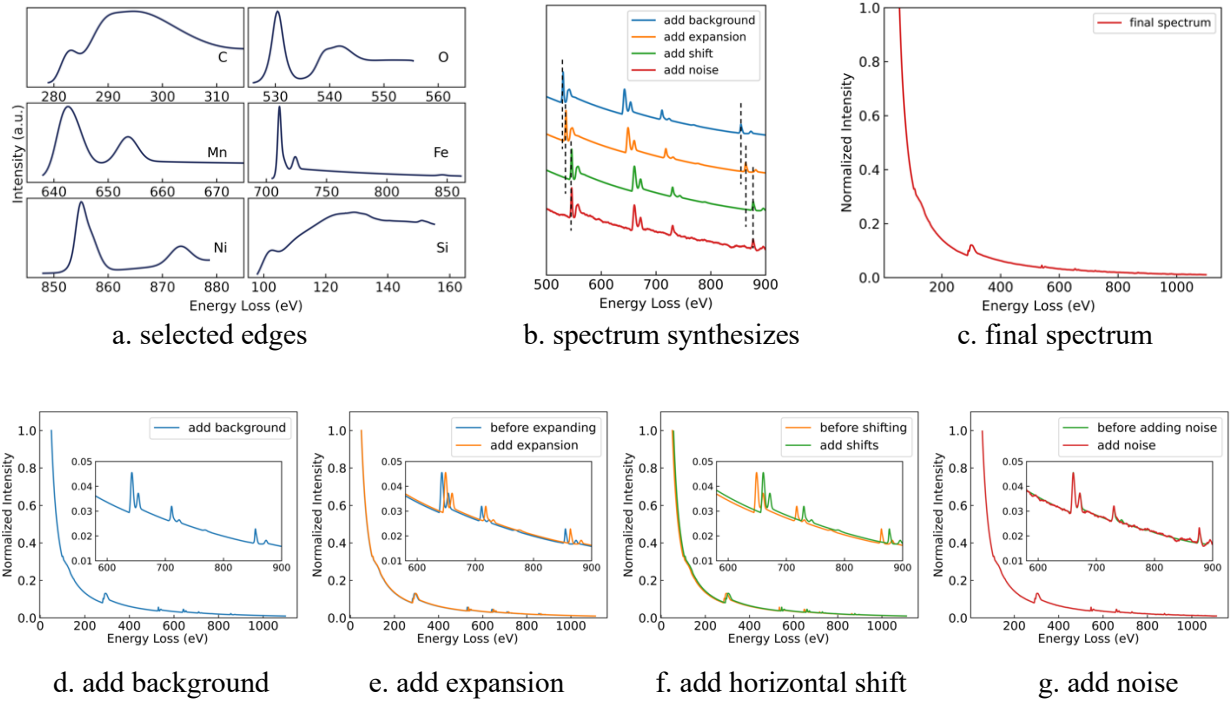

Fig. 3 Spectrum Synthesize Steps and Process

With the jump ratio randomly selected from the corresponding range for the element, the extended edge can be added to the background, which is shown in Fig. 3 (d).

In the real case, distortions and translational shifts of the spectra can be introduced due to multiple reasons including the hysteresis of the prism and the magnifying lenses, the spectrometer alignment, stray magnetic fields, and even temperature, etc. As a result, the spectra collected at different times can vary for the same instrument. To simulate these phenomena, we allocate a maximum of 1.5% of the contraction or expansion ratio to the modeled spectrum. A random shift of the spectrum is also implemented to the spectrum with a maximum shift of 15 eV. Fig. 3 (b) and (e) show the expanded spectrum. The spectrum that is further shifted is shown in Fig. 3 (b) and (f).

The last step of the spectrum synthesis is to add correct noise to it. Firstly, Gaussian noise with a mean of 0 and a standard deviation of 0.0012 is generated. Afterward, the Savitzky-Golay filter is applied to smooth the noise to low-frequency noise, where the polynomial order of the filter is set to 1, and the frame length is an odd positive integer in the range of 30 to 60. Fig. 4 shows the raw

Gaussian noise and the filtered noise. The filtered low-frequency noise is also called fixed-pattern noise. It is frequently seen in experimental EELS spectra primarily due to the non-uniform gain on the scintillator. To augment the noise level, a random coefficient C that is in the range from 1 to 2 is further assigned to the noise. Fig. 3 (g) presents the spectrum with noise added, which is also the input spectrum for our deep learning network. Table 2 summarizes the data augmentation operations in the process of spectrum synthesis. We generated 25,000 spectra in total, where the percentage of the spectra containing 1 to 6 edges are 20 %, 20 %, 20 %, 15 %, 15 %, and 10 %, respectively.

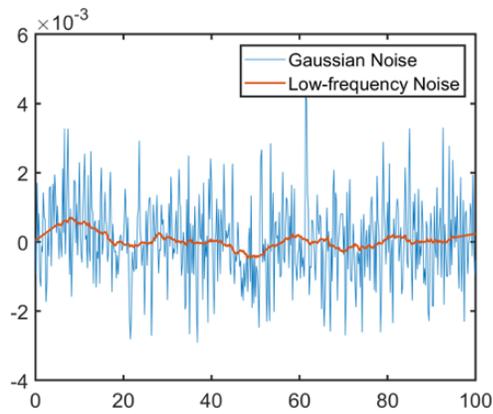

Fig. 4 Gaussian Noise and Filtered Noise

Table 2 Augmentation Operations Summary

| Operation | Randomness Range |
|---|---|
| Edge Select | Types of Elements: 1 ~ 6 |
| PSFs Convolve | $FWHM_g$: 0.1 ~ 1.6 <br> $HWHM_l$: 0.1 ~ 0.6 |
| Edge Extend | Power $r'$: -3.8 ~ - 1.8 |
| Background | $r$: Gaussian (1.88, 0.5) > 0.1 |
| Jump Ratio | Range shown in Table 1 |
| Contraction/Expansion | -1.5 % ~ 1.5% |
| Horizontal Shift | -15 eV ~ 15 eV |

| Gaussian Noise | Gaussian (0, 0.0012) |
| --- | --- |
| Savitzky-Golay Filter | Frame length: 30 ~ 60 odd number |
| | Coefficient $C$: 1.0 ~ 2.0 |

**Network Construction**

*CNN-BiLSTM Network*  CNN is a type of network developed for conducting image recognition and has been widely implemented in image classification. A typical CNN structure is constructed by connecting a few convolutional layers and dense layers (fully connected layers). The convolutional layers can extract the features from the input data and the dense layers classify the features to achieve classification. The recognition of the ionization edges on a spectrum is also a type of image recognition task. In the meanwhile, the onset ionization edge carries the information of the target element. The edge's energy position relative to others plays a critical role in the classification process. However, a pure CNN structure is insensitive to the positional/sequential information. Given that the long short-term memory network (LSTM) enables the network to memorize the information in sequence by selectively keeping the important data and deleting less important data to pass to the next layer, we proposed a CNN-BiLSTM combined network for ionization edge recognition and classification in the core loss region of the spectrum. Fig. 5 (a) shows the architecture of our CNN-BiLSTM network, where the BiLSTM network is plotted in an unwrapped way. The real LSTM cell is described in Fig. 5 (b), and is constructed with 100 hidden units. In our case, the input layer carries the information of a spectrum that is a 1D vector (spectrum of energy loss range from 90 eV to 1000 eV with interval 0.2 eV).  A filter of size 11 is implemented in each convolutional layer. Batch normalization operation[66] is carried out following each convolutional layer, and 20 % of node dropout regularization is utilized after each dense layer to avoid overfitting. Activation functions, ReLU and LeakyReLU of scale 0.01, are implemented to connect different layers as shown in the Fig. 5 (a). The network is built and trained using MATLAB. The training parameter settings are listed in Table 3. 80 % of the spectra in the database are randomly picked for the training set, and the remaining 20 % are for the validation set.

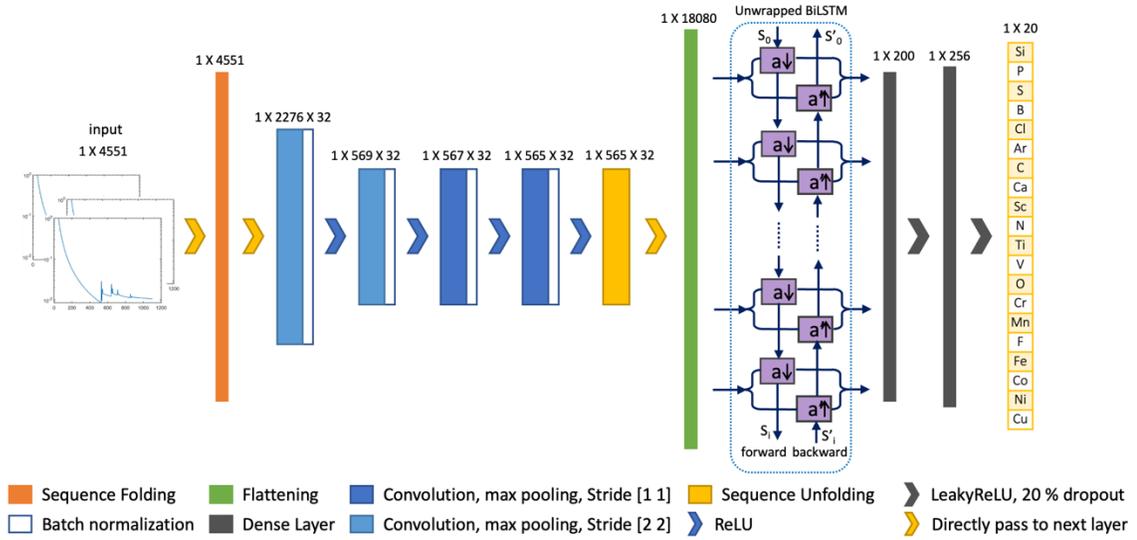

(a) CNN-BiLSTM

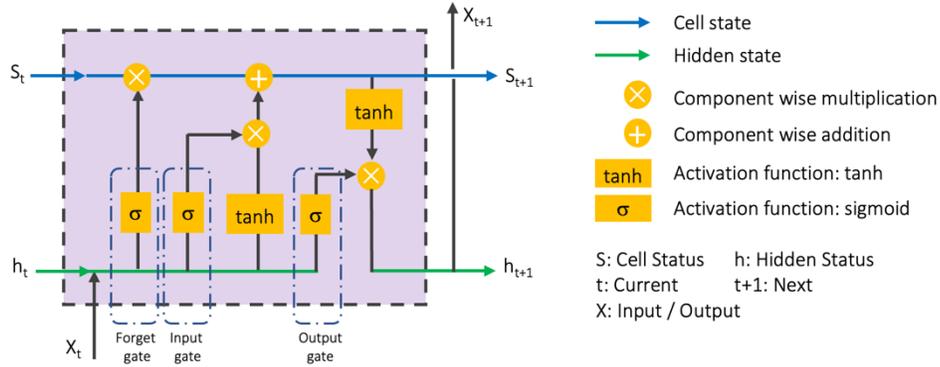

(b) LSTM Cell $\vec{a}$

Fig. 5 CNN-BiLSTM Network Architecture

Table 3 Network Training Setting

| Options | Setting |
| --- | --- |
| ExecutionEnvironment | GPU |
| InitialLearnRate | 0.001 |
| GradientThreshold | 1 |
| Shuffle | every-epoch |
| ValidationPatience | Inf |
| ValidationFrequency | 50 |
| Verbose | true |

## RESULTS AND DISCUSSION

### Prediction Validation

The performance of the proposed CNN-BiLSTM network is validated using multiple metrics.

*F1-score*  To evaluate the performance of the network per spectrum, we calculate the F1-score for each predicted spectrum, which is defined as the harmonic mean of precision and recall. We use $NE_c$, $NE_p$, and $NE_r$ to represent the number of correctly predicted edges (true positive), the total number of predicted edges (predicted positive), and the number of real edges (actual positive) on the spectrum, respectively. Precision, recall, and F1-score can be calculated by using the following equations:

$$precision = \frac{NE_c}{NE_p}$$

$$recall = \frac{NE_c}{NE_r}$$

$$F1 = \frac{2}{recall^{-1} + precision^{-1}}$$

*Exact Accuracy*  We use exact accuracy (ACC) to check how our network can accurately recognize all the edges on a spectrum, i.e., the ratio of the number of spectra accurately predicted ($N_{exa}$) to the total number of spectra tested ($N$). This value also reflects the percentage of the spectra whose F1-score equals to 1.

$$Acc = \frac{N_{exa}}{N}$$

*Recognition Accuracy*     The following equation shows the format of the recognition accuracy. $N_{rec}$ represents the number of spectra that the network can recognize all the edges that appear on them. In other words, the percentage of the spectra whose recall is 1. This matric suggests how sensitive the network can detect the available edges.

$$Rec = \frac{N_{rec}}{N}$$

*Network Validation*     Besides the synthesized validation dataset, we also use 160 real experimental spectra to examine the performance of our network.

Fig. 6(a) shows the values of validation metrics for both the real spectra dataset and the validation dataset (The precision, recall, and F1-score are the averaged values). Fig. 6 (b) presents the boxplots of the F1-scores of the two datasets, where only some outliers have F1-scores less than 1, which means that our network is highly accurate in recognizing the edges. The average F1-scores of the real spectra dataset and the validation dataset are as high as 0.99 and 0.98, respectively. The comparable high F1-scores of the two datasets further indicate that the network is well-trained without overfitting. The average precision of the real spectra dataset and the validation dataset are 0.98 and 0.99, respectively. The average recall of the real spectra dataset and the validation dataset are 0.99 and 0.97, respectively. The high values of the precision and recall indicate that false positive and false negative edges barely appear in a prediction.

The exact accuracy and the recognition accuracy are two metrics describing the percentages of the desired spectra. For the real dataset, the exact accuracy reaches 94.9 %. And the network sensitively detects all the edges presented on the spectra, which can be reflected by the recognition accuracy of 98.9 %. In other words, only around 1 % of the predicted spectra have edges that failed to be detected by our network. The exact accuracy and recognition accuracy of the validation set is relatively lower than those of the real spectra dataset, but they are still around 90 %. This phenomenon could be caused by the higher diversity of the synthesized spectra, where we stretched the parameters' ranges during data augmentation to create more extreme cases.

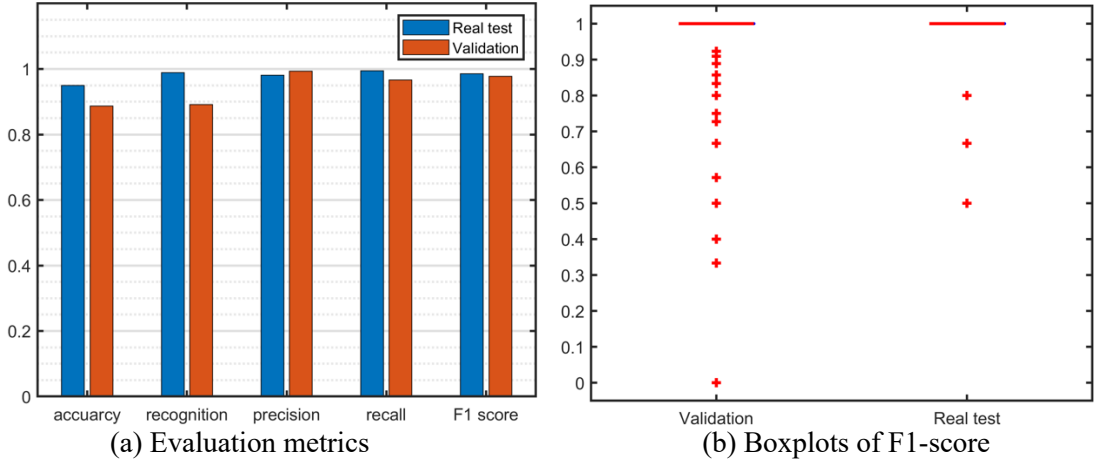

(a) Evaluation metrics

(b) Boxplots of F1-score

Fig. 6 Performance of the well-trained CNN-BiLSTM network

**Evaluation of Model's Sensitivity**

*Peak Signal Noise Ratio (PSNR)* To further evaluate the sensitivity of the model in detecting the ionization edges for a specific element, we introduce the concept of Peak Signal Noise Ratio, which is the signal noise ratio at the peak of the edge. It is expressed by the ratio of the signal at the edge peak off the inverse power law background (shown in Fig. 2 as $I_p$) to the standard deviation of the noise (*sd*). In other words,

$$PSNR = \frac{I_p}{sd}$$

To make the value of PSNR for different elements comparable, we further synthesize one spectrum for each element with the value of the power *r* for the background set to be the same constant 1.88. We added different levels of the Gaussian noise (white noise) to the spectra to create more extreme situations to examine the sensitivity of our network. The standard deviation of the noise is set to be one-tenth of the minimum value of the background:

$$std = \frac{\min_{50\sim1100} J_n(E)}{10}$$

*Sensitivity Evaluation*  To quantify the sensitivity of our network, we searched for the minimum value of *PSNR* for each element that the network can correctly recognize by varying the scale of the peak. Correspondingly, the value of the jump ratio can be captured simultaneously. The results are listed in Table 4.

It is worth noting that the detectable jump ratios in Table 4 are much lower than the minimum value of 0.15 we set for building the training dataset. This result illustrates that our network can manage the situation when the jump ratios of the edges are lower than those commonly observed in experiments. In the meanwhile, the noise is not smoothed by the Savitzky-Golay filter at this stage. Still, our network can sensitively extract valuable information from the noisy spectra. Higher detectable *PSNRs* for light elements are expected because these edges tend to appear in a relatively high-intensity lower energy loss region. Referring to the corresponding low jump ratios, the network performs well in recognizing these light elements' ionization edges.

Table 4 Detectable PSNR and Jump Ratio

| Element | PSNR | **Jump Ratio** | Element | PSNR | **Jump Ratio** |
|---------|------|----------------|---------|------|----------------|
| B | 23.00 | **0.077** | Mn | 3.50 | **0.090** |
| C | 10.40 | **0.071** | Fe | 3.68 | **0.116** |
| N | 5.87 | **0.068** | Co | 3.52 | **0.130** |
| O | 4.51 | **0.088** | Ni | 3.05 | **0.126** |
| F | 3.45 | **0.099** | Cu | 3.16 | **0.158** |
| Ca | 7.69 | **0.075** | Si | 56.29 | **0.059** |
| Sc | 6.29 | **0.078** | P | 50.26 | **0.086** |
| Ti | 5.36 | **0.083** | S | 19.26 | **0.049** |
| V | 3.99 | **0.070** | Cl | 16.82 | **0.062** |
| Cr | 3.85 | **0.083** | Ar | 7.94 | **0.040** |

**CONCLUSION**

In this work, we built a CNN-BiLSTM network for automatically detecting the ionization edges on EELS spectra. We also created a forward model for synthesizing EELS spectra, which allows us to create a database containing 25000 spectra with ground-truth labels for training and testing the network. The network is examined by using both the synthesized validation dataset and the real spectra dataset collected from experiments. Multiple validation matrices are calculated to evaluate the performance of the network regarding the two datasets, including exact accuracy, recognition accuracy, average precision, average recall, and average F1-score. The sensitivity of the network is explored by testing the lowest detectable PSNR and detectable jump ratio.

The average F1-score, recall, and precision for the real spectra dataset agree well with those for the validation dataset, which suggests that, for one thing, the synthesized spectra are comparable to real spectra, for the other, the network is well-trained without overfitting. The high values of these three metrics also illustrate the good performance of the network in respect to the low false positive or false negative rate. We find that for 98.9 % of the real spectra, our network is able to recognize all the edges. For 94.9 % of them, our network can accurately recognize all the edges without false positive edges, which means that only 4.0 % of the spectra have false positive edges, and 1.1 % of the spectra have false negative edges. What's more, the low value of the detectable PSNRs and jump ratios further demonstrate the high sensitivity of our network.

Taken together, our CNN-BiLSTM network can automatically recognized the core-loss edges on a raw spectrum with high accuracy. It is also safe to be applied to EELS spectra where the SNR and jump ratio are extremely low. The success of the network in automatically recognizing core-loss edges on a raw spectrum further makes it possible to the automation of more advanced core-loss edge analysis.